# Closed-Form Solution of the Unit Normal Loss Integral in Two-Dimensions

Tae Yoon Lee, MSc[1]; Paul Gustafson, PhD[2]; Mohsen Sadatsafavi, MD, PhD[1]; for the Canadian Respiratory Research Network

1. Respiratory Evaluation Sciences Program, Collaboration for Outcomes Research and Evaluation, Faculty of Pharmaceutical Sciences, University of British Columbia, Vancouver, Canada
2. Department of Statistics, University of British Columbia, Vancouver, Canada

**Corresponding author:**
> Tae Yoon Lee, MSc
> PhD Candidate, Faculty of Pharmaceutical Sciences
> University of British Columbia
> https://resp.core.ubc.ca/team/Harry_Lee
> Email: dlxodbs@student.ubc.ca

**Running head:** Two-dimensional unit normal loss integral

**Word count:** 1,472

Financial support for this study was provided in part by a grant from Genome Canada/Genome British Columbia (274CHI), and the Canadian Respiratory Research Network (CRRN). CRRN is supported by grants from the Canadian Institutes of Health Research - Institute of Circulatory and Respiratory Health; Canadian Lung Association/Canadian Thoracic Society; British Columbia Lung Association; and Industry Partners Boehringer-Ingelheim Canada Ltd, AstraZeneca Canada Inc., and Novartis Canada Ltd. Funding for training of graduate students and new investigators within the Network was supported by the above funding Sponsors and as well by GlaxoSmithKline Inc. The funding Sponsors had no role in the study design, data collection and analysis, or preparation of the manuscript. The funding agreement ensured the authors' independence in designing the study, interpreting the data, writing, and publishing the report.

**ABSTRACT**

In Value of Information (VoI) analysis, the unit normal loss integral (UNLI) frequently emerges as a solution for the computation of various VoI metrics. However, one limitation of the UNLI has been that its closed-form solution is available for only one dimension, and thus can be used for comparisons involving only two strategies (where it is applied to the scalar incremental net benefit). We derived a closed-form solution for the two-dimensional UNLI, enabling closed-form VoI calculations for three strategies. We verified the accuracy of this method via simulation studies. A case study based on a three-arm clinical trial was used as an example. VoI methods based on the closed-form solutions for the UNLI can now be extended to three-decision comparisons, taking a fraction of a second to compute and not being subject to Monte Carlo error. An R implementation of this method is provided as part of the *predtools* package (https://github.com/resplab/predtools/).

**Introduction**

Value of Information (VoI) analysis is a set of concepts and methods rooted in decision theory that quantifies the expected utility loss due to uncertainty associated with decisions, or the expected utility gain by conducting further research.(1) VoI has been applied in decision analysis across areas including health technology assessment,(2) environmental risk analysis,(3) and clinical prediction modeling.(4) Numerical or Bayesian calculations underlying VoI are often carried out using Monte Carlo (MC) simulations, either through repeated sampling of uncertain input parameters in model-based evaluations or via bootstrapping in data-driven analyses.(5)

The Unit Normal Loss Integral (UNLI), first discussed by Raiffa and Schlaifer in 1960s, has emerged in various ways in VoI analysis.(6) An overview of VoI metrics and UNLI-based computations is provided by Willson.(7) The UNLI is closely related to the mean of the truncated normal distribution. The exact definition of the UNLI has varied slightly in different publications.(7–9) Here we use the following definition. Suppose Y has a normal distribution with mean µ and variance $\sigma^2$. With $\phi$ and $\Phi$ denoting the probability density and cumulative distribution functions of the standard normal distribution, the UNLI can be defined as:

$$UNLI = E(\max(Y,0)) = \int_0^\infty \frac{y}{\sigma} \phi\left(\frac{y-\mu}{\sigma}\right) dy = \mu\left[1 - \Phi\left(-\frac{\mu}{\sigma}\right)\right] + \sigma\phi\left(-\frac{\mu}{\sigma}\right).$$

A typical instance in which the UNLI is used in VoI is the computation of the Expected Value of Perfect Information (EVPI).(6,10) EVPI is the expected gain in net benefit when uncertainty in the evidence underlying the decision is completely resolved.(7) When comparing two strategies (e.g., use of a new medication versus continuing with standard of care for treating a disease), the outcome of a probabilistic decision analysis can be summarized as a distribution of the incremental net benefit (INB) between the two strategies. If this quantity has a normal distribution, then the EVPI can be expressed as a closed-form solution using the one-dimensional UNLI, as outlined above.(6,10) This approach for EVPI calculation is applicable to both model-based and data-driven evaluations (with the latter, the normality assumption is supported by the central limit theorem). The UNLI method has been extended to other VoI metrics, such as Expected Value of Partial Perfect Information(11) and Expected Value of Sample Information.(12) Such solutions are computationally feasible and free from MC error.

However, in many practical decision analyses, there are more than two decisions that are compared, making the closed-form UNLI not readily applicable. Approximate methods have been suggested for more than two strategies. For example, Jalal et al use the UNLI as an approximate solution to multiple comparisons by segmenting the joint probability space of input parameters into adjacent pieces to turn the problem into a sum of one-dimensional evaluations.(9)

To the best of our knowledge, no closed-form expression for the UNLI has been proposed for higher dimensions. In this work, we derived a closed-form solution for the UNLI for two dimensions, enabling the extension of this method to comparisons of three strategies. We performed a simulation study to verify the numerical accuracy of this method, and showed its utility in a case study involving EVPI calculation for a three-arm clinical trial.

**Closed-form solution**

Suppose we have three strategies of interest, with one labelled as the reference strategy (the choice of which strategy being designated as the reference has no bearing on the computation). The INBs of the two alternative strategies compared with the reference strategy are denoted by $Y_1$ and $Y_2$. Further suppose that our knowledge of $(Y_1, Y_2)$ can be expressed as a bivariate normal distribution (BVN) with mean $(\mu_1, \mu_2)$, variance $(\sigma_1^2, \sigma_2^2)$, and correlation coefficient $\rho$. The expected NB under perfect information is the expectation of $max(Y_1, Y_2, 0)$.

Following the derivations provided in the Supplementary Material – Section 1, we arrive at the closed-form equation:

$$E(\max(Y_1, Y_2, 0)) = u_{1,2} + v_{1,2} + u_{2,1} + v_{2,1},$$

where

$$u_{i,j} = \mu_i \left[ \mathbf{1}\{(\sigma_i - \rho\sigma_j) > 0\} + \Phi\left(\frac{-\sigma_i\mu_j + \rho\sigma_j\mu_i}{\sigma_i\sigma_j\sqrt{(1-\rho^2)}}\right) \mathbf{1}\{(\sigma_i - \rho\sigma_j) = 0\} \right] -$$

$$\Phi\left(\frac{-\sigma_i\mu_j + \rho\sigma_j\mu_i}{\sigma_i\sigma_j\sqrt{(1-\rho^2)}}\right) \left(-\sigma_i\phi\left(\frac{-\mu_i}{\sigma_i}\right) + \mu_i\Phi\left(\frac{-\mu_i}{\sigma_i}\right)\right),$$

and with $\alpha_{i,j} = \frac{\sigma_i \mu_j - \rho \sigma_j \mu_i}{\sigma_i - \rho \sigma_j}$, $\beta_{i,j} = \frac{\sigma_i \sigma_j \sqrt{(1-\rho^2)}}{\sigma_i - \rho \sigma_j}$, $a_{1ij} = \frac{\mu_i - \alpha_{i,j}}{|\beta_{i,j}|}$, $b_{1ij} = \frac{\sigma_i}{|\beta_{i,j}|}$, $a_{2ij} = \frac{\alpha_{i,j} - \mu_i}{\sigma_i}$, $b_{2ij} = \frac{|\beta_{i,j}|}{\sigma_i}$, and $t_{kij} = \sqrt{(1 + b_{kij}^2)}$;

$$v_{i,j} = \mu_i sgn(\beta_{i,j}) \left[ \Phi\left( \frac{-a_{1ij}/b_{1ij}}{\sqrt{(1 + (1/b_{1ij})^2)}} \right) - \Phi_2\left( \frac{-a_{1ij}}{t_{1ij}}, \frac{-\alpha_{i,j}}{|\beta_{i,j}|}, -\frac{1}{t_{1ij}} \right) \right]$$

$$- sgn(\beta_{i,j}) \frac{\sigma_i}{t_{2ij}} \phi\left( \frac{a_{2ij}}{t_{2ij}} \right) \left( 1 - \Phi\left( \frac{-t_{2ij}\alpha_{i,j}}{|\beta_{i,j}|} + \frac{a_{2ij}b_{2ij}}{t_{2ij}} \right) \right).$$

where $\Phi_2(x_1, x_2, \rho)$ is the cumulative density function of the standard bivariate normal with the upper limits $x_1$ and $x_2$, and correlation coefficient $\rho$.

**Simulation study**

We conducted a simulation study to confirm the correctness of the closed-form solution by comparing it with large-scale MC integrations (N=100,000). We examined 252 permutations of the parameters of the bivariate distributions characterized by a factorial design for the following variables: $\mu_1, \mu_2 = \{-2, 0, 2\}$, $\sigma_1^2, \sigma_2^2 = \{1, 3\}$, and $\rho = \{-0.75, -0.50, ..., 0.50, 0.75\}$. Results show that the difference between the closed-form and MC solutions falls within the range of the MC error (Supplementary Material – Section 2).

**Case study**

The Optimal Therapy of Chronic Obstructive Pulmonary Disease (COPD) was a parallel-arm clinical trial of three inhaler therapies for patients with COPD. In this three-arm study, 449 patients with COPD were randomized to receive single-inhaler (N=145), double-inhaler (N=156), or triple-inhaler (N=148) therapies.(14) The trial duration was 12 months. The study collected data, including monthly cost diaries, and functional scores measured by St. George's Respiratory Questionnaire at baseline and four follow-up visits (at 4, 20, 36, and 52 months).

The net benefit calculations in this study closely follow the methods used in a previously published data-driven economic evaluation of this study.(15) The functional scores were converted to EQ5D utilities using validated algorithms.(16) With the Quality-Adjusted Life Years (QALY) as the health outcome of interest, the single inhaler strategy dominated the double inhaler strategy, and the incremental cost effectiveness ratio of the triple inhaler therapy versus the single inhaler therapy was $(4,042-2,678)/(0.7217-0.7092) = $243,180 per QALY gained (costs are in 2006 Canadian dollars). As such, the single-inhaler therapy was the optimal strategy at the willingness-to-pay (WTP) value of $50,000 per QALY. However, a bootstrap-based probabilistic sensitivity analysis demonstrated uncertainty in the results: the single-inhaler therapy was the optimal strategy in 80% of the bootstraps.

We calculated the EVPI for this evaluation based on such individual-level data from the trial, comparing the bootstrap-based approach(5) with the proposed UNLI method. At a given WTP value, the net benefit was calculated as QALY multiplied by WTP minus total costs for each patient. Then for each arm $j$, we calculated the average NB, $\widehat{NB}_j$. 7% of costs and utility values were missing. Similar to the approach used in the original evaluation, we imputed the missing values using multiple imputation by chained equation with predictive mean matching.(17) For the bootstrap-based approach, imputation was embedded within bootstrapping, such that each iteration of the MC simulation involved one imputation and generation of a single bootstrapped sample. This was repeated 1,000 times.

For the UNLI method, we generated 10 imputed datasets and pooled the mean and covariance matrix estimates of $\widehat{INB}s$ using the Rubin's rule.(18) Next, taking the single-inhaler therapy as the reference, the INBs of double- and triple-inhaler therapies were parameterized as a bivariate normal distribution, as specified previously. For example, at WTP of $50,000/QALY, the pooled parameter values were $\mu_1 = -4,734$, $\mu_2 = -2,668$, $\sigma_1 = 4,678$, $\sigma_2 = 4,645$, and ρ=0.50, where subscripts 1 and 2 indicate the INB of double- and triple-inhaler therapies, respectively, compared with single-inhaler therapy.

EVPI values were similar between the closed-form UNLI and bootstrap methods with the mean relative absolute error of 3% (*Figure 1*). At WTP of $50,000/QALY, the EVPI values were

$1,019 for the closed-form UNLI method and $976 (MC standard error: 61) for the bootstrap method. At WTP of $100,000/QALY, the corresponding values were $1,570 and $1,543 (MC standard error: 107).

<<Figure 1>>

## Discussion

We have proposed a closed-form solution for the two-dimensional version of the UNLI, enabling VoI calculations for decision analyses with three strategies. The UNLI method takes just a fraction of a second to compute and is not subject to the MC error inherent in simulation-based methods. The R code for this method is provided in the mu_max_trunc_bvn function in the *predtools* R package (https://github.com/resplab/predtools).

This approach can be useful for data-driven and model-based evaluations with three comparators, and can be particularly helpful with computationally expensive evaluations. In a case study, we compared this method with the conventional bootstrap-based approach for EVPI computations for a three-arm clinical trial. The one-dimensional UNLI method has been expanded to other VoI metrics for evaluations of two strategies, including the Expected Value of Partial Perfect Information and the Expected Value of Sample Information.(7,9) Those algorithms can be feasibly modified to accommodate decision analyses with three strategies using the two-dimensional UNLI method. Such modified algorithms would obviate the need for an approximation that is currently required.

This approach requires the normality assumption on the joint distribution of the incremental net benefits of two alternative strategies with respect to a default strategy. In data-driven evaluations, the Central Limit Theorem provides a general justification for the normality assumption with sufficient sample sizes.(7) For model-based evaluation, the appropriateness of the normality assumption must be ascertained, as has been demonstrated in the previous applications of the UNLI method.(19)

A natural question is whether the UNLI method can be extended to higher dimensions. We believe the answer is no. The $m$-dimensional UNLI is closely related to the expression for the maximum of $m$-variate Gaussian random variables. Arellano-Valle and Genton show that for m>2, closed-form solutions for this maximum are generally unavailable (Corollary 4).(20)

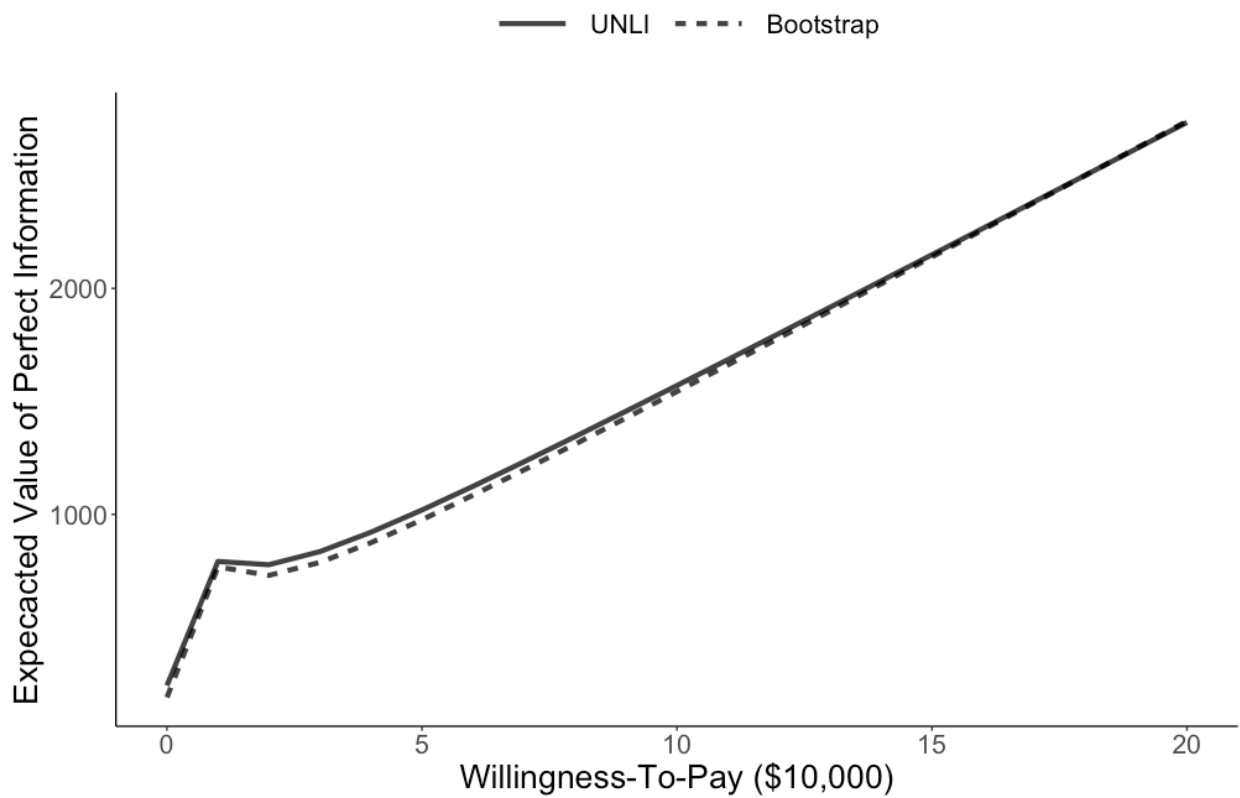

**Figure 1**. Expected Value of Perfect Information based on the closed-form Unit Normal Loss Integral (UNLI) method (solid) and bootstrap method (dashed).

# Supplementary Materials

## 1. Derivation of the closed-form solution for two-dimension UNLI

Suppose we have two strategies of interest in comparison to a reference strategy, with corresponding two scalar incremental net benefits, denoted by $Y_1$ and $Y_2$. Further suppose that $Y = (Y_1, Y_2)$ follow a bivariate normal distribution with mean $(\mu_1, \mu_2)$, variance $(\sigma_1^2, \sigma_2^2)$, and correlation coefficient $\rho$. Our target quantity is the expectation of $max(Y, 0)$:

$$E(\max(Y, 0)) = E(Y\mathbf{1}\{Y > 0\} + 0\mathbf{1}\{Y \leq 0\})$$
$$= E(Y\mathbf{1}\{Y > 0\})$$
$$= \int_{-\infty}^{\infty} y\, f_Y(y)\mathbf{1}\{Y > 0\}dy$$
$$= \int_{-\infty}^{\infty} y\, f_{Y_1}(-y)\mathbf{1}\{Y > 0\}dy + \int_{-\infty}^{\infty} y\, f_{Y_2}(-y)\mathbf{1}\{Y > 0\}dy,$$

where $\mathbf{1}(\cdot)$ is the indicator function, and the last equation follows from the decomposition of the bivariate normal distribution,[1] with $\phi$ and $\Phi$ denoting the probability density and cumulative distribution functions of the standard normal distribution:

$$f_{X_i}(x_i) = \frac{1}{\sigma_i}\phi\left(\frac{x_i + \mu_i}{\sigma_i}\right)\Phi\left(\frac{\rho(x_i + \mu_i)}{\sigma_i\sqrt{(1-\rho^2)}} - \frac{x_i + \mu_j}{\sigma_j\sqrt{(1-\rho^2)}}\right), i \neq j, i, j = 1, 2.$$

By symmetry, it suffices to compute the first term:

$$\int_{-\infty}^{\infty} y\, f_{Y_1}(-y)\mathbf{1}\{Y > 0\}dy = \int_0^{\infty} y\frac{1}{\sigma_1}\phi\left(\frac{y - \mu_1}{\sigma_1}\right)\Phi\left(\frac{\sigma_1(y - \mu_2) - \sigma_2\rho(y - \mu_1)}{\sigma_1\sigma_2\sqrt{(1-\rho^2)}}\right)dy.$$

We apply the integration by parts technique with

$$u = \Phi\left(\frac{\sigma_1(y - \mu_2) - \sigma_2\rho(y - \mu_1)}{\sigma_1\sigma_2\sqrt{(1-\rho^2)}}\right) = \Phi\left(\frac{y(\sigma_1 - \rho\sigma_2) - \sigma_1\mu_2 + \rho\sigma_2\mu_1}{\sigma_1\sigma_2\sqrt{(1-\rho^2)}}\right)$$

and

$$dv = y\frac{1}{\sigma_1}\phi\left(\frac{(y - \mu_1)}{\sigma_1}\right)dy.$$

Then we have

$$du = \frac{(\sigma_1 - \rho\sigma_2)}{\sigma_1\sigma_2\sqrt{(1-\rho^2)}}\phi\left(\frac{y(\sigma_1 - \rho\sigma_2) - \sigma_1\mu_2 + \rho\sigma_2\mu_1}{\sigma_1\sigma_2\sqrt{(1-\rho^2)}}\right)dy$$

and

$$v = \int_0^\infty y \frac{1}{\sigma_1} \phi\left(\frac{(y-\mu_1)}{\sigma_1}\right) dy = -\sigma_1 \phi\left(\frac{y-\mu_1}{\sigma_1}\right) + \mu_1 \Phi\left(\frac{y-\mu_1}{\sigma_1}\right).$$

The first component is straightforward:

$$[uv|_0^\infty = \mu_1 \left[ \mathbf{1}\{(\sigma_1 - \rho\sigma_2) > 0\} + \Phi\left(\frac{-\sigma_1\mu_2+\rho\sigma_2\mu_1}{\sigma_1\sigma_2\sqrt{(1-\rho^2)}}\right) \mathbf{1}\{(\sigma_1 - \rho\sigma_2) = 0\}\right] -$$

$$\Phi\left(\frac{-\sigma_1\mu_2+\rho\sigma_2\mu_1}{\sigma_1\sigma_2\sqrt{(1-\rho^2)}}\right)\left(-\sigma_1\phi\left(\frac{-\mu_1}{\sigma_1}\right) + \mu_1\Phi\left(\frac{-\mu_1}{\sigma_1}\right)\right).$$

We compute term-by-term for the second component:

$$\int_0^\infty v\, du = 1/\beta \int_0^\infty \left(\mu_1 \Phi\left(\frac{y-\mu_1}{\sigma_1}\right) - \sigma_1 \phi\left(\frac{y-\mu_1}{\sigma_1}\right)\right) \phi\left(\frac{y-\alpha}{\beta}\right) dy =: T_1 + T_2.$$

Let $\alpha = \frac{\sigma_1\mu_2 - \rho\sigma_2\mu_1}{\sigma_1 - \rho\sigma_2}$, $\beta = \frac{\sigma_1\sigma_2\sqrt{(1-\rho^2)}}{\sigma_1 - \rho\sigma_2}$, $sgn(\cdot)$ be the sign function, and $sBVN(w_1, w_2, \rho)$ be the cumulative density of the standard bivariate normal with the upper limits $w_1, w_2$ and correlation coefficient $\rho$.

We use the table of integrals by Owen,[2] and we have two cases depending on the sign of $\beta$. We solve the case for $\beta > 0$.

$$T_1 = \frac{\mu_1}{\beta} \int_0^\infty \Phi\left(\frac{y - \mu_1}{\sigma_1}\right) \phi\left(\frac{y - \alpha}{\beta}\right) dy$$

use change of variable with x=(y-α)/β

$$= \frac{\mu_1}{\beta} \int_{-\alpha/\beta}^\infty \phi(x) \Phi\left(\frac{\beta x + \alpha - \mu_1}{\sigma_1}\right) (\beta dx)$$

let $a = (\mu_1 - \alpha)/\beta$ and $b = \sigma_1/\beta$

$$= \mu_1 \int_{-\alpha/\beta}^\infty \phi(x) \Phi\bigl((x-a)/b\bigr) dx$$

$$= \mu_1 \left[\int_{-\infty}^\infty \phi(x)\Phi\bigl((x-a)/b\bigr)dx - \int_{-\infty}^{-\alpha/\beta} \phi(x)\Phi\bigl((x-a)/b\bigr)dx\right]$$

$$= \mu_1 \left[\Phi\bigl((-a/b)/(1+(1/b)^2)^{0.5}\bigr) - sBVN\bigl(-a/\sqrt{(1+b^2)}, -\alpha/\beta, -1/\sqrt{(1+b^2)}\bigr)\right]$$

$$T_2 = \frac{-\sigma_1}{\beta} \int_0^\infty \phi\left(\frac{y - \mu_1}{\sigma_1}\right) \phi\left(\frac{y - \alpha}{\beta}\right) dy$$

use change of variable $x = (y - \alpha)/\beta$

$$= \frac{-\sigma_1}{\beta} \int_{-\alpha/\beta}^\infty \phi\left(\frac{x\beta + \alpha - \mu_1}{\sigma_1}\right) \phi(x)(\beta dx)$$

let $a = (\alpha - \mu_1)/\sigma_1, b = \beta/\sigma_1, t = \sqrt{1 + b^2}$

$$= -\sigma_1 \int_{-\alpha/\beta}^\infty \phi(a + bx)\phi(x)dx$$

$$= -\sigma_1 \left[\frac{1}{t}\phi(a/t)\Phi(tx + ab/t)\right|_{-\alpha/\beta}^\infty$$

$$= \frac{-\sigma_1}{t}\phi(a/t)\left(1 - \Phi(-t\alpha/\beta + ab/t)\right)$$

We carry out similar calculations for $\beta < 0$. Then we arrive at the following general expression with $a_1 = \frac{\mu_1 - \alpha}{|\beta|} b_1 = \frac{\sigma_1}{|\beta|}, a_2 = \frac{\alpha - \mu_1}{\sigma_1}, b_2 = \frac{|\beta|}{\sigma_1}$, and $t_i = \sqrt{(1 + b_i^2)}$,

$$\int_0^\infty v\, du = sgn(\beta)\mu_1\left[\Phi\left((-a_1/b_1)/(1 + (1/b_1)^2)^{0.5}\right) - sBVN(-a_1/t_1, -\alpha/\beta, -1/t_1)\right] +$$

$$-sgn(\beta)\frac{\sigma_1}{t_2}\phi(a_2/t_2)\left(1 - \Phi(-t_2\alpha/\beta + a_2 b_2/t_2)\right).$$

## 2. Simulation results

**Table 1.** Comparison of the closed-form and Monte Carlo (N=100,000) solutions for 252 different bivariate distributions.

| $\mu_1$ | $\mu_2$ | $\sigma_1^2$ | $\sigma_2^2$ | $\rho$ | Closed-form | Monte Carlo | $\mu_1$ | $\mu_2$ | $\sigma_1^2$ | $\sigma_2^2$ | $\rho$ | Closed-form | Monte-Carlo |
|---|---|---|---|---|---|---|---|---|---|---|---|---|---|
| -2 | -2 | 1 | 1 | -0.75 | 0.017 | 0.017 | -2 | -2 | 1 | 3 | 0 | 0.114 | 0.113 |
| 0 | -2 | 1 | 1 | -0.75 | 0.407 | 0.405 | 0 | -2 | 1 | 3 | 0 | 0.477 | 0.476 |
| 2 | -2 | 1 | 1 | -0.75 | 2.014 | 2.016 | 2 | -2 | 1 | 3 | 0 | 2.024 | 2.026 |
| -2 | 0 | 1 | 1 | -0.75 | 0.407 | 0.409 | -2 | 0 | 1 | 3 | 0 | 0.696 | 0.694 |
| 0 | 0 | 1 | 1 | -0.75 | 0.772 | 0.772 | 0 | 0 | 1 | 3 | 0 | 0.944 | 0.947 |
| 2 | 0 | 1 | 1 | -0.75 | 2.136 | 2.139 | 2 | 0 | 1 | 3 | 0 | 2.170 | 2.170 |
| -2 | 2 | 1 | 1 | -0.75 | 2.014 | 2.015 | -2 | 2 | 1 | 3 | 0 | 2.108 | 2.107 |
| 0 | 2 | 1 | 1 | -0.75 | 2.136 | 2.137 | 0 | 2 | 1 | 3 | 0 | 2.196 | 2.196 |
| 2 | 2 | 1 | 1 | -0.75 | 2.746 | 2.746 | 2 | 2 | 1 | 3 | 0 | 2.799 | 2.801 |
| -2 | -2 | 3 | 1 | -0.75 | 0.115 | 0.116 | -2 | -2 | 3 | 3 | 0 | 0.206 | 0.205 |
| 0 | -2 | 3 | 1 | -0.75 | 0.699 | 0.701 | 0 | -2 | 3 | 3 | 0 | 0.761 | 0.758 |
| 2 | -2 | 3 | 1 | -0.75 | 2.114 | 2.109 | 2 | -2 | 3 | 3 | 0 | 2.133 | 2.137 |
| -2 | 0 | 3 | 1 | -0.75 | 0.505 | 0.504 | -2 | 0 | 3 | 3 | 0 | 0.761 | 0.756 |
| 0 | 0 | 3 | 1 | -0.75 | 1.057 | 1.055 | 0 | 0 | 3 | 3 | 0 | 1.180 | 1.173 |
| 2 | 0 | 3 | 1 | -0.75 | 2.321 | 2.324 | 2 | 0 | 3 | 3 | 0 | 2.323 | 2.322 |
| -2 | 2 | 3 | 1 | -0.75 | 2.067 | 2.070 | -2 | 2 | 3 | 3 | 0 | 2.133 | 2.138 |
| 0 | 2 | 3 | 1 | -0.75 | 2.321 | 2.322 | 0 | 2 | 3 | 3 | 0 | 2.323 | 2.327 |
| 2 | 2 | 3 | 1 | -0.75 | 3.025 | 3.019 | 2 | 2 | 3 | 3 | 0 | 2.984 | 2.982 |
| -2 | -2 | 1 | 3 | -0.75 | 0.115 | 0.113 | -2 | -2 | 1 | 1 | 0.25 | 0.017 | 0.016 |
| 0 | -2 | 1 | 3 | -0.75 | 0.505 | 0.507 | 0 | -2 | 1 | 1 | 0.25 | 0.402 | 0.399 |
| 2 | -2 | 1 | 3 | -0.75 | 2.067 | 2.064 | 2 | -2 | 1 | 1 | 0.25 | 2.009 | 2.009 |
| -2 | 0 | 1 | 3 | -0.75 | 0.699 | 0.702 | -2 | 0 | 1 | 1 | 0.25 | 0.402 | 0.401 |
| 0 | 0 | 1 | 3 | -0.75 | 1.057 | 1.053 | 0 | 0 | 1 | 1 | 0.25 | 0.643 | 0.643 |
| 2 | 0 | 1 | 3 | -0.75 | 2.321 | 2.320 | 2 | 0 | 1 | 1 | 0.25 | 2.032 | 2.034 |
| -2 | 2 | 1 | 3 | -0.75 | 2.114 | 2.117 | -2 | 2 | 1 | 1 | 0.25 | 2.009 | 2.006 |
| 0 | 2 | 1 | 3 | -0.75 | 2.321 | 2.325 | 0 | 2 | 1 | 1 | 0.25 | 2.032 | 2.032 |
| 2 | 2 | 1 | 3 | -0.75 | 3.025 | 3.020 | 2 | 2 | 1 | 1 | 0.25 | 2.489 | 2.484 |
| -2 | -2 | 3 | 3 | -0.75 | 0.213 | 0.212 | -2 | -2 | 3 | 1 | 0.25 | 0.113 | 0.114 |
| 0 | -2 | 3 | 3 | -0.75 | 0.796 | 0.803 | 0 | -2 | 3 | 1 | 0.25 | 0.694 | 0.689 |
| 2 | -2 | 3 | 3 | -0.75 | 2.191 | 2.188 | 2 | -2 | 3 | 1 | 0.25 | 2.107 | 2.110 |
| -2 | 0 | 3 | 3 | -0.75 | 0.796 | 0.800 | -2 | 0 | 3 | 1 | 0.25 | 0.460 | 0.458 |
| 0 | 0 | 3 | 3 | -0.75 | 1.337 | 1.335 | 0 | 0 | 3 | 1 | 0.25 | 0.898 | 0.904 |

| | | | | | | | | | | | | |
|---|---|---|---|---|---|---|---|---|---|---|---|---|
| 2 | 0 | 3 | 3 | -0.75 | 2.533 | 2.529 | 2 | 0 | 3 | 1 | 0.25 | 2.160 | 2.155 |
| -2 | 2 | 3 | 3 | -0.75 | 2.191 | 2.190 | -2 | 2 | 3 | 1 | 0.25 | 2.015 | 2.012 |
| 0 | 2 | 3 | 3 | -0.75 | 2.533 | 2.529 | 0 | 2 | 3 | 1 | 0.25 | 2.120 | 2.120 |
| 2 | 2 | 3 | 3 | -0.75 | 3.293 | 3.290 | 2 | 2 | 3 | 1 | 0.25 | 2.708 | 2.706 |
| -2 | -2 | 1 | 1 | -0.5 | 0.017 | 0.017 | -2 | -2 | 1 | 3 | 0.25 | 0.113 | 0.113 |
| 0 | -2 | 1 | 1 | -0.5 | 0.407 | 0.406 | 0 | -2 | 1 | 3 | 0.25 | 0.460 | 0.462 |
| 2 | -2 | 1 | 1 | -0.5 | 2.012 | 2.014 | 2 | -2 | 1 | 3 | 0.25 | 2.015 | 2.018 |
| -2 | 0 | 1 | 1 | -0.5 | 0.407 | 0.410 | -2 | 0 | 1 | 3 | 0.25 | 0.694 | 0.697 |
| 0 | 0 | 1 | 1 | -0.5 | 0.744 | 0.744 | 0 | 0 | 1 | 3 | 0.25 | 0.898 | 0.900 |
| 2 | 0 | 1 | 1 | -0.5 | 2.107 | 2.104 | 2 | 0 | 1 | 3 | 0.25 | 2.120 | 2.125 |
| -2 | 2 | 1 | 1 | -0.5 | 2.012 | 2.019 | -2 | 2 | 1 | 3 | 0.25 | 2.107 | 2.108 |
| 0 | 2 | 1 | 1 | -0.5 | 2.107 | 2.110 | 0 | 2 | 1 | 3 | 0.25 | 2.160 | 2.156 |
| 2 | 2 | 1 | 1 | -0.5 | 2.691 | 2.691 | 2 | 2 | 1 | 3 | 0.25 | 2.708 | 2.712 |
| -2 | -2 | 3 | 1 | -0.5 | 0.115 | 0.114 | -2 | -2 | 3 | 3 | 0.25 | 0.198 | 0.199 |
| 0 | -2 | 3 | 1 | -0.5 | 0.699 | 0.699 | 0 | -2 | 3 | 3 | 0.25 | 0.741 | 0.740 |
| 2 | -2 | 3 | 1 | -0.5 | 2.112 | 2.115 | 2 | -2 | 3 | 3 | 0.25 | 2.119 | 2.127 |
| -2 | 0 | 3 | 1 | -0.5 | 0.499 | 0.496 | -2 | 0 | 3 | 3 | 0.25 | 0.741 | 0.741 |
| 0 | 0 | 3 | 1 | -0.5 | 1.023 | 1.026 | 0 | 0 | 3 | 3 | 0.25 | 1.114 | 1.113 |
| 2 | 0 | 3 | 1 | -0.5 | 2.277 | 2.276 | 2 | 0 | 3 | 3 | 0.25 | 2.253 | 2.246 |
| -2 | 2 | 3 | 1 | -0.5 | 2.050 | 2.052 | -2 | 2 | 3 | 3 | 0.25 | 2.119 | 2.113 |
| 0 | 2 | 3 | 1 | -0.5 | 2.271 | 2.273 | 0 | 2 | 3 | 3 | 0.25 | 2.253 | 2.257 |
| 2 | 2 | 3 | 1 | -0.5 | 2.955 | 2.956 | 2 | 2 | 3 | 3 | 0.25 | 2.862 | 2.864 |
| -2 | -2 | 1 | 3 | -0.5 | 0.115 | 0.113 | -2 | -2 | 1 | 1 | 0.5 | 0.016 | 0.016 |
| 0 | -2 | 1 | 3 | -0.5 | 0.499 | 0.502 | 0 | -2 | 1 | 1 | 0.5 | 0.400 | 0.400 |
| 2 | -2 | 1 | 3 | -0.5 | 2.050 | 2.047 | 2 | -2 | 1 | 1 | 0.5 | 2.008 | 2.007 |
| -2 | 0 | 1 | 3 | -0.5 | 0.699 | 0.692 | -2 | 0 | 1 | 1 | 0.5 | 0.400 | 0.403 |
| 0 | 0 | 1 | 3 | -0.5 | 1.023 | 1.026 | 0 | 0 | 1 | 1 | 0.5 | 0.598 | 0.599 |
| 2 | 0 | 1 | 3 | -0.5 | 2.271 | 2.272 | 2 | 0 | 1 | 1 | 0.5 | 2.016 | 2.018 |
| -2 | 2 | 1 | 3 | -0.5 | 2.112 | 2.106 | -2 | 2 | 1 | 1 | 0.5 | 2.008 | 2.013 |
| 0 | 2 | 1 | 3 | -0.5 | 2.277 | 2.271 | 0 | 2 | 1 | 1 | 0.5 | 2.016 | 2.018 |
| 2 | 2 | 1 | 3 | -0.5 | 2.955 | 2.948 | 2 | 2 | 1 | 1 | 0.5 | 2.400 | 2.402 |
| -2 | -2 | 3 | 3 | -0.5 | 0.213 | 0.214 | -2 | -2 | 3 | 1 | 0.5 | 0.111 | 0.112 |
| 0 | -2 | 3 | 3 | -0.5 | 0.789 | 0.792 | 0 | -2 | 3 | 1 | 0.5 | 0.692 | 0.697 |
| 2 | -2 | 3 | 3 | -0.5 | 2.170 | 2.175 | 2 | -2 | 3 | 1 | 0.5 | 2.107 | 2.105 |
| -2 | 0 | 3 | 3 | -0.5 | 0.789 | 0.787 | -2 | 0 | 3 | 1 | 0.5 | 0.440 | 0.441 |
| 0 | 0 | 3 | 3 | -0.5 | 1.289 | 1.290 | 0 | 0 | 3 | 1 | 0.5 | 0.845 | 0.839 |
| 2 | 0 | 3 | 3 | -0.5 | 2.462 | 2.464 | 2 | 0 | 3 | 1 | 0.5 | 2.130 | 2.130 |
| -2 | 2 | 3 | 3 | -0.5 | 2.170 | 2.171 | -2 | 2 | 3 | 1 | 0.5 | 2.010 | 2.003 |
| 0 | 2 | 3 | 3 | -0.5 | 2.462 | 2.462 | 0 | 2 | 3 | 1 | 0.5 | 2.072 | 2.077 |

| | | | | | | | | | | | | |
|---|---|---|---|---|---|---|---|---|---|---|---|---|
| 2 | 2 | 3 | 3 | -0.5 | 3.197 | 3.195 | 2 | 2 | 3 | 1 | 0.5 | 2.605 | 2.608 |
| -2 | -2 | 1 | 1 | -0.25 | 0.017 | 0.017 | -2 | -2 | 1 | 3 | 0.5 | 0.111 | 0.112 |
| 0 | -2 | 1 | 1 | -0.25 | 0.406 | 0.410 | 0 | -2 | 1 | 3 | 0.5 | 0.440 | 0.440 |
| 2 | -2 | 1 | 1 | -0.25 | 2.010 | 2.012 | 2 | -2 | 1 | 3 | 0.5 | 2.010 | 2.010 |
| -2 | 0 | 1 | 1 | -0.25 | 0.406 | 0.405 | -2 | 0 | 1 | 3 | 0.5 | 0.692 | 0.693 |
| 0 | 0 | 1 | 1 | -0.25 | 0.714 | 0.714 | 0 | 0 | 1 | 3 | 0.5 | 0.845 | 0.841 |
| 2 | 0 | 1 | 1 | -0.25 | 2.079 | 2.080 | 2 | 0 | 1 | 3 | 0.5 | 2.072 | 2.074 |
| -2 | 2 | 1 | 1 | -0.25 | 2.010 | 2.006 | -2 | 2 | 1 | 3 | 0.5 | 2.107 | 2.110 |
| 0 | 2 | 1 | 1 | -0.25 | 2.079 | 2.078 | 0 | 2 | 1 | 3 | 0.5 | 2.130 | 2.125 |
| 2 | 2 | 1 | 1 | -0.25 | 2.631 | 2.627 | 2 | 2 | 1 | 3 | 0.5 | 2.605 | 2.609 |
| -2 | -2 | 3 | 1 | -0.25 | 0.115 | 0.115 | -2 | -2 | 3 | 3 | 0.5 | 0.185 | 0.185 |
| 0 | -2 | 3 | 1 | -0.25 | 0.698 | 0.697 | 0 | -2 | 3 | 3 | 0.5 | 0.719 | 0.723 |
| 2 | -2 | 3 | 1 | -0.25 | 2.110 | 2.112 | 2 | -2 | 3 | 3 | 0.5 | 2.110 | 2.114 |
| -2 | 0 | 3 | 1 | -0.25 | 0.490 | 0.489 | -2 | 0 | 3 | 3 | 0.5 | 0.719 | 0.720 |
| 0 | 0 | 3 | 1 | -0.25 | 0.985 | 0.980 | 0 | 0 | 3 | 3 | 0.5 | 1.036 | 1.033 |
| 2 | 0 | 3 | 1 | -0.25 | 2.235 | 2.234 | 2 | 0 | 3 | 3 | 0.5 | 2.185 | 2.185 |
| -2 | 2 | 3 | 1 | -0.25 | 2.036 | 2.033 | -2 | 2 | 3 | 3 | 0.5 | 2.110 | 2.104 |
| 0 | 2 | 3 | 1 | -0.25 | 2.221 | 2.224 | 0 | 2 | 3 | 3 | 0.5 | 2.185 | 2.187 |
| 2 | 2 | 3 | 1 | -0.25 | 2.880 | 2.887 | 2 | 2 | 3 | 3 | 0.5 | 2.719 | 2.707 |
| -2 | -2 | 1 | 3 | -0.25 | 0.115 | 0.114 | -2 | -2 | 1 | 1 | 0.75 | 0.014 | 0.015 |
| 0 | -2 | 1 | 3 | -0.25 | 0.490 | 0.489 | 0 | -2 | 1 | 1 | 0.75 | 0.399 | 0.400 |
| 2 | -2 | 1 | 3 | -0.25 | 2.036 | 2.033 | 2 | -2 | 1 | 1 | 0.75 | 2.008 | 2.006 |
| -2 | 0 | 1 | 3 | -0.25 | 0.698 | 0.692 | -2 | 0 | 1 | 1 | 0.75 | 0.399 | 0.399 |
| 0 | 0 | 1 | 3 | -0.25 | 0.985 | 0.978 | 0 | 0 | 1 | 1 | 0.75 | 0.540 | 0.541 |
| 2 | 0 | 1 | 3 | -0.25 | 2.221 | 2.221 | 2 | 0 | 1 | 1 | 0.75 | 2.009 | 2.006 |
| -2 | 2 | 1 | 3 | -0.25 | 2.110 | 2.111 | -2 | 2 | 1 | 1 | 0.75 | 2.008 | 2.004 |
| 0 | 2 | 1 | 3 | -0.25 | 2.235 | 2.232 | 0 | 2 | 1 | 1 | 0.75 | 2.009 | 2.007 |
| 2 | 2 | 1 | 3 | -0.25 | 2.880 | 2.881 | 2 | 2 | 1 | 1 | 0.75 | 2.285 | 2.283 |
| -2 | -2 | 3 | 3 | -0.25 | 0.211 | 0.211 | -2 | -2 | 3 | 1 | 0.75 | 0.108 | 0.106 |
| 0 | -2 | 3 | 3 | -0.25 | 0.777 | 0.777 | 0 | -2 | 3 | 1 | 0.75 | 0.691 | 0.693 |
| 2 | -2 | 3 | 3 | -0.25 | 2.150 | 2.150 | 2 | -2 | 3 | 1 | 0.75 | 2.107 | 2.105 |
| -2 | 0 | 3 | 3 | -0.25 | 0.777 | 0.778 | -2 | 0 | 3 | 1 | 0.75 | 0.418 | 0.419 |
| 0 | 0 | 3 | 3 | -0.25 | 1.237 | 1.240 | 0 | 0 | 3 | 1 | 0.75 | 0.781 | 0.778 |
| 2 | 0 | 3 | 3 | -0.25 | 2.392 | 2.394 | 2 | 0 | 3 | 1 | 0.75 | 2.110 | 2.117 |
| -2 | 2 | 3 | 3 | -0.25 | 2.150 | 2.158 | -2 | 2 | 3 | 1 | 0.75 | 2.009 | 2.007 |
| 0 | 2 | 3 | 3 | -0.25 | 2.392 | 2.394 | 0 | 2 | 3 | 1 | 0.75 | 2.031 | 2.034 |
| 2 | 2 | 3 | 3 | -0.25 | 3.095 | 3.096 | 2 | 2 | 3 | 1 | 0.75 | 2.479 | 2.480 |
| -2 | -2 | 1 | 1 | 0 | 0.017 | 0.017 | -2 | -2 | 1 | 3 | 0.75 | 0.108 | 0.108 |
| 0 | -2 | 1 | 1 | 0 | 0.404 | 0.402 | 0 | -2 | 1 | 3 | 0.75 | 0.418 | 0.420 |
| 2 | -2 | 1 | 1 | 0 | 2.009 | 2.008 | 2 | -2 | 1 | 3 | 0.75 | 2.009 | 2.008 |
| -2 | 0 | 1 | 1 | 0 | 0.404 | 0.405 | -2 | 0 | 1 | 3 | 0.75 | 0.691 | 0.689 |
| 0 | 0 | 1 | 1 | 0 | 0.681 | 0.679 | 0 | 0 | 1 | 3 | 0.75 | 0.781 | 0.781 |
| 2 | 0 | 1 | 1 | 0 | 2.053 | 2.048 | 2 | 0 | 1 | 3 | 0.75 | 2.031 | 2.031 |
| -2 | 2 | 1 | 1 | 0 | 2.009 | 2.010 | -2 | 2 | 1 | 3 | 0.75 | 2.107 | 2.102 |
| 0 | 2 | 1 | 1 | 0 | 2.053 | 2.057 | 0 | 2 | 1 | 3 | 0.75 | 2.110 | 2.116 |
| 2 | 2 | 1 | 1 | 0 | 2.564 | 2.562 | 2 | 2 | 1 | 3 | 0.75 | 2.479 | 2.483 |
| -2 | -2 | 3 | 1 | 0 | 0.114 | 0.116 | -2 | -2 | 3 | 3 | 0.75 | 0.165 | 0.163 |
| 0 | -2 | 3 | 1 | 0 | 0.696 | 0.692 | 0 | -2 | 3 | 3 | 0.75 | 0.698 | 0.697 |
| 2 | -2 | 3 | 1 | 0 | 2.108 | 2.110 | 2 | -2 | 3 | 3 | 0.75 | 2.107 | 2.099 |
| -2 | 0 | 3 | 1 | 0 | 0.477 | 0.474 | -2 | 0 | 3 | 3 | 0.75 | 0.698 | 0.701 |
| 0 | 0 | 3 | 1 | 0 | 0.944 | 0.940 | 0 | 0 | 3 | 3 | 0.75 | 0.935 | 0.934 |
| 2 | 0 | 3 | 1 | 0 | 2.196 | 2.197 | 2 | 0 | 3 | 3 | 0.75 | 2.126 | 2.128 |
| -2 | 2 | 3 | 1 | 0 | 2.024 | 2.026 | -2 | 2 | 3 | 3 | 0.75 | 2.107 | 2.106 |
| 0 | 2 | 3 | 1 | 0 | 2.170 | 2.170 | 0 | 2 | 3 | 3 | 0.75 | 2.126 | 2.122 |
| 2 | 2 | 3 | 1 | 0 | 2.799 | 2.798 | 2 | 2 | 3 | 3 | 0.75 | 2.537 | 2.537 |